\documentclass[preprint,showpacs,preprintnumbers,amsmath,amssymb]{revtex4}

\usepackage{titlesec}
\usepackage{graphicx,epsfig,epstopdf}
\usepackage{bm}
\usepackage{cleveref}
\usepackage{amssymb,amsmath,amsfonts,latexsym}
\usepackage{dcolumn}
\usepackage{bm}
\usepackage[utf8]{inputenc}
\newcommand{\nn}{\nonumber\\}

\begin{document}

\title{Decay ${\mathit{D} \to} {{\mathit K}^{(*)}}{{\mathit \ell}^{+}}{{\mathit \nu}_{{{{\mathit \ell}}}}}$ in covariant quark model}

\author{N. R. Soni}
\email{nrsoni-apphy@msubaroda.ac.in}
\affiliation{Applied Physics Department, Faculty of Technology and Engineering, The Maharaja Sayajirao University of Baroda, Vadodara 390001, Gujarat, India}

\author{J. N. Pandya}
\email{jnpandya-apphy@msubaroda.ac.in}
\affiliation{Applied Physics Department, Faculty of Technology and Engineering, The Maharaja Sayajirao University of Baroda, Vadodara 390001, Gujarat, India}

\date{\today}

\begin{abstract}
We study the leptonic and semileptonic $D$-meson decays (${{\mathit D} \to} {{\mathit \ell}^{+}}{{\mathit \nu}_{{{{\mathit \ell}}}}}$ and  ${\mathit{D} \to} {{\mathit K}^{(*)}}{{\mathit \ell}^{+}}{{\mathit \nu}_{{{{\mathit \ell}}}}}$) in the framework of covariant quark model with built-in infrared confinement. We compute the required form factors in the entire kinematical momentum transfer region. The calculated form factors are used to evaluate the branching fractions of these transitions. We determine the following ratios of the partial widths: $\Gamma ({{\mathit D}^{0}} \rightarrow {{\mathit K}^{-}}{{\mathit e}^{+}}{{\mathit \nu}_{{e}}})/\Gamma({{\mathit D}^{+}} \rightarrow {{\overline{\mathit K}}^{0}}{{\mathit e}^{+}}{{\mathit \nu}_{{e}}}) = 1.02$, $\Gamma({{\mathit D}^{0}} \rightarrow {{\mathit K}^{-}}{{\mathit \mu}^{+}}{{\mathit \nu}_{{\mu}}})/\Gamma({{\mathit D}^{+}} \rightarrow {{\overline{\mathit K}}^{0}}{{\mathit \mu}^{+}}{{\mathit \nu}_{{\mu}}}) = 0.99$ and $\Gamma({{\mathit D}^{+}} \rightarrow {{\overline{\mathit K}}^{0}}{{\mathit \mu}^{+}}{{\mathit \nu}_{{\mu}}}) / \Gamma({{\mathit D}^{+}} \rightarrow {{\overline{\mathit K}}^{0}}{{\mathit e}^{+}}{{\mathit \nu}_{{e}}}) = 0.97$ which are in close resemblance with the iso-spin invariance and experimental results.
\end{abstract}

\maketitle

\section{\label{sec:introduction}Introduction}
The semileptonic decays involve strong as well as weak interactions. The extraction of Cabibbo-Kobayashi-Maskawa (CKM) matrix elements from these exclusive decays can be parameterized by form factor calculations.
As $|V_{cd}|$ and $|V_{cs}|$ are constrained by CKM unitarity, the calculation of semileptonic decays of $D$-mesons can also be an important test to look for new physics.
The decay $D \rightarrow K^{(*)} \ell^+ \nu_{\ell}$ provides accurate determination of $|V_{cs}|$. Thus, the theoretical prediction for the form factors and their $q^2$-dependence need to be tested. A comprehensive review of experimental and theoretical challenges in study of hadronic decays of $D$ and $D_s$ mesons along with required experimental and theoretical tools \cite{Anders:2012} provide motivation to look into semileptonic decays in this paper.

Recently, BESIII \cite{ablikim:2017lks,ablikim:2016sqt,ablikim:2015,Ablikim:2015qgt} and \textit{BABAR} \cite{Lees:2015} collaborations have reported precise and improved measurements on semileptonic form factors and branching fractions on decays of $D \rightarrow K \ell^+ \nu_{\ell}$ and $D \rightarrow \pi \ell^+ \nu_{\ell}$. A brief review of the earlier work and present experimental status of $D$-meson decays are given in \cite{Amhis:2016}.
Also there are variety of theoretical models available in the literature for the computation of hadronic form factors. One of the oldest model is based on the quark model known as ISGW model for CP violation in semileptonic $B$ meson decays based on the nonrelativistic constituent quark picture \cite{Isgur:1989}. The advanced version (ISGW2 model \cite{Scora:1995}) includes the heavy quark symmetry and has been used for semileptonic decays of $B_{(s)}$, $D_{(s)}$ and $B_c$ mesons. The form factors are also calculated in Lattice Quantum Chromodynamics (LQCD) \cite{Aubin:2004,Bernard:2009,Na:2010,AlHaydari:2010,Na:2011,Koponen:2013},  light-cone sum rules (LCSR) \cite{Khodjamirian:2000,Khodjamirian:2007,Khodjamirian:2009} and LCSR with heavy quark effective theory \cite{Wang:2003}. The form factor calculations from LCSR provide good results at low ($q^2\simeq 0$) and high ($q^2\simeq q_{max}^2$) momentum transfers. The form factors have also been calculated for the process $D \to K \ell \nu_{\ell}$ in the entire momentum transfer range \cite{Koponen:2013} using the LQCD. Also recently the Flavour Lattice Averaging Group (FLAG) have reported the latest lattice results for determination of CKM matrices within the standard model\cite{Aoki:2016}.

The form factors of $D, B \to P, V, S$ transitions with $P$, $V$ and $S$ corresponding to pseudoscalar, vector and scalar meson respectively have been evaluated in the light front quark model (LFQM) \cite{Verma:2011}. The form factors for $D \to P, V$ are also computed in the framework of chiral quark model ($\chi$QM) \cite{Palmer:2013} as well in the phenomenological model based on heavy meson chiral theory (HM$\chi$T) \cite{Fajfer:2004,Fajfer:2005}. The form factors of $B_{(s)},D_{(s)} \to \pi, K, \eta$ have been evaluated in three flavor hard pion chiral perturbation theory \cite{Bijnens:2011}. The form factors for $D \to \pi e^+ \nu_e$ have been computed in the framework of ``charm-changing current'' \cite{Ananthanarayan:2011}. The authors of \cite{Rong:2014,Fang:2014} have determined the form factors $f_+^{K(\pi)}$  by globally analysing the available measurements of branching fractions for $D \to K(\pi) e^+ \nu_e$.
The vector form factors for $D \to K \ell \nu_\ell$ were also parameterized in \cite{DescotesGenon:2008}.  The evaluation of transition form factors and decays of $B_{(s)},D_{(s)} \to f_0(980),K_0^*(1430) \ell \nu_\ell$ has been done in \cite{Yang:2005,Bediaga:2003} from QCD sum rules. The computation of differential branching fractions for $D_{(s)} \to (P, V, S) \ell \nu_\ell$ was also performed using chiral unitary approach \cite{Sekihara:2015,Xie:2014}, generalized linear sigma model \cite{Fariborz:2014,Fariborz:2011} and sum rules \cite{Wang:2010}. Various decay properties of $D_{(s)}$ and $B_{(s)}$ are also studied in the formalism of semi-relativistic \cite{Kher:2017,Devlani:2013,Devlani:2012,Devlani:2011} and relativistic  \cite{Shah:2016,Shah:2014,Shah:2014d} potential models.

In this paper, we employ the covariant constituent quark model (CQM) with built-in infrared confinement \cite{Efimov:1988,Efimov:1993,Ivanov:1999,Ivanov:2000,Faessler:2002,Branz:2010} to compute the leptonic and semileptonic decays. The form factors of these transitions are expressed through only few universal functions. One of the key feature of CQM is access to the entire physical range of momentum transfer. Our aim is to perform independent calculations of these decays including  $q^2$ behavior of the transition form factors, leptonic decay constants of $D$ and $K$ mesons and ratios of branching fractions for the decay  ${\mathit{D} \to} {{\mathit K}^{(*)}}{{\mathit \ell}^{+}}{{\mathit \nu}_{{{{\mathit \ell}}}}}$ and ${\mathit{D} \to} {\mathit \pi}{{\mathit \ell}^{+}}{{\mathit \nu}_{{{{\mathit \ell}}}}}$.

This paper is organised as follows. After a brief introduction of the semileptonic $D$-meson decays in Sec. \ref{sec:introduction}, in Sec. \ref{sec:model} we introduce the theoretical framework of CQM and also discuss the method of extracting the model parameters. In Sec. \ref{sec:form_factors}, we give the definition of the form factors for the decays  ${\mathit{D} \to} {{\mathit K}^{(*)}}{{\mathit \ell}^{+}}{{\mathit \nu}_{{{{\mathit \ell}}}}}$.
In Sec. \ref{sec:results} for numerical results, we first compute the leptonic branching fractions of $D^+$-meson. Next we give numerical results of the form factors. We also parameterize  the form factors using  double pole approximation. From the form factors, we compute the differential branching fraction for the ${\mathit{D} \to} {{\mathit K}^{(*)}}{{\mathit \ell}^{+}}{{\mathit \nu}_{{{{\mathit \ell}}}}}$, with $\ell = e$  and $\mu$ and the branching fractions. We also calculate the forward-backward asymmetry and convexity parameters. We compare our results with available experimental, lattice and other theoretical results. Finally, we summarize present work in Sec. \ref{sec:summary}.
\section{Model}
\label{sec:model}
The CQM is an effective quantum field approach \cite{Efimov:1988,Efimov:1993,Ivanov:1999,Ivanov:2000,Faessler:2002,Branz:2010} for hadronic interactions based on an effective Lagrangian of hadrons interacting with their constituent quarks. In this paper, we restrict ourselves to weak decays of $D$-mesons only. The interaction Lagrangian describing the coupling of meson $M(q_1 \bar{q}_2)$ to the constituent quarks $q_1$ and $\bar{q}_2$ in the framework of CQM is given by
\begin{eqnarray}\label{eq:int_lagrange}
\mathcal{L}_{int} &  = & g_M M(x) \int dx_1 dx_2 F_M(x;x_1,x_2) \bar{q}_2(x_2) \Gamma_M q_1(x_1) \nn & + &  H.c.
\end{eqnarray}
where $\Gamma_M$ is the Dirac matrix and projects onto the spin quantum number of relevant mesonic field $M(x)$. $g_M$ is the coupling constant and $F_M$ is the vertex function that is related to the scalar part of the Bethe-Salpeter amplitude. $F_M$ also characterizes the finite size of the mesons. We choose the vertex function that satisfies the Lorentz invariance of the Lagrangian Eq. (\ref{eq:int_lagrange}),
\begin{equation}
F_M(x,x_1,x_2) = \delta \left(x - \sum_{i=1}^2 w_i x_i \right)\Phi_M \left((x_1 - x_2)^2\right)
\end{equation}
with $\Phi_M$ is the correlation function of two constituent quarks with masses $m_{q_1}$ and $m_{q_2}$ and $w_{q_i} = m_{q_i}/ (m_{q_1} + m_{q_2})$ such that ${\mathit{w_1}} + {\mathit {w_2}} = 1$.
We choose Gaussian function for vertex function as
\begin{equation} \label{eq:gaussian}
\tilde{\Phi}_M(-p^2) = \exp \ (p^2/\Lambda_M^2)
\end{equation}
with the parameter $\Lambda_M$ characterized by the finite size of the meson.
In the Euclidian space, we can write $p^2 = -p^2_E$, so that the vertex function has the appropriate falloff behaviour so as to remove the ultraviolet divergence in the loop integral.

We use the compositeness conditions \cite{Salam:1962,Weinberg:1963} to determine the coupling strength $g_M$ in Eq. (\ref{eq:mass_operator}) that requires the renormalization constant $Z_M$ for the bare state to composite mesonic state $M(x)$ set to zero, i.e.,
\begin{equation}
Z_M = 1 -  \tilde\Pi'_M(m^2_M) = 0,
\label{eq:Z=0}
\end{equation}
where $\tilde\Pi'_M$ is the derivative of meson mass operator and $Z_M$ is the wave function renormalization constant of the meson $M$. Here, $Z_M^{1/2}$ is the matrix element between the physical state and the corresponding bare state. The above condition guarantees that the physical state does not contain any bare quark state i.e. bound state. The constituents are virtual and are introduced to realize the interaction and as a result the physical state turns dressed and its mass and wave function are renormalized.

\begin{figure}[htbp]
\includegraphics{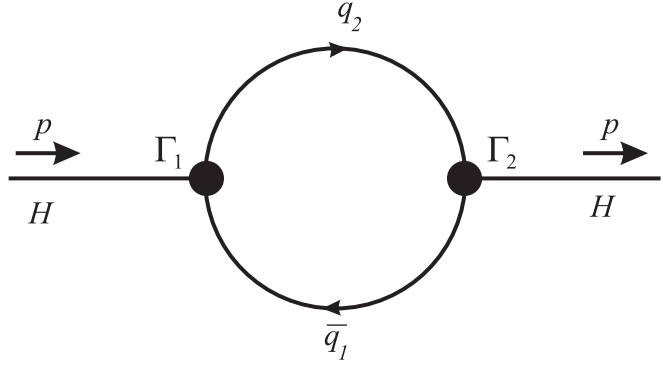}
\caption{Diagram describing meson mass operator.}
\label{fig:mass}
\end{figure}

The meson mass operator Fig. \ref{fig:mass} for any meson is defined as
\begin{eqnarray}\label{eq:mass_operator}
\tilde\Pi_M(p^2) &=& N_c g_{M}^{2}
\int\frac{d^4k}{(2\pi)^4i} \tilde\Phi^2_M(-k^2) \nn & \times &
\mathrm{tr}\Big(\Gamma_1 S_1(k+w_1 p)\Gamma_2 S_2(k-w_2 p)\Big)
\end{eqnarray}
where $N_c=3$ is the number of colors. $\Gamma_1$, $\Gamma_2$ are the Dirac matrices and for scalar, vector and pseudoscalar mesons, we choose the gamma matrices accordingly. $S's$ are the quark propagator and we use the free fermion propagator for the constituent quark. For the computation of loop integral in Eq. (\ref{eq:mass_operator}), we write the quark propagator in terms of Fock-Schwinger representation as
\begin{eqnarray}\label{eq:Fock}
S_q (k+ p) &=& \frac{1}{ m_q-\not\! k-  \not\! p }
=  \frac{m_q + \not\! k +  \not\! p}{m^2_q - (k+ p)^2}
\nn
&=& (m_q + \not\! k +  \not\! p)\int\limits_0^\infty \!\!d\alpha\,
e^{-\alpha [m_q^2-(k+ p)^2]}.
\end{eqnarray}
where $k$ is the loop momentum and $p$ is the external momentum. The use of Fock-Schwinger representation allows to do the tensor integral in an efficient way since the loop momenta can be converted into the derivative of exponential function \cite{Branz:2010}. All the necessary trace evaluation and loop integrals are done in FORM \cite{Vermaseren:2008}. For the remaining integral over the Fock-Schwinger parameters $0 \leq \alpha_i \leq \infty$, we use an additional integration converting the Fock-Schwinger parameters into a simplex. The transformation reads \cite{Feynman:1949}
\begin{eqnarray}\label{eq:mass_int}
&&\prod\limits_{i=1}^n\int\limits_0^{\infty}
\!\! d\alpha_i f(\alpha_1,\ldots,\alpha_n) \\
&&=\int\limits_0^{\infty} \!\! dtt^{n-1}
\prod\limits_{i=1}^n \int\!\!d\alpha_i
\delta\left(1-\sum\limits_{i=1}^n\alpha_i\right)
  f(t\alpha_1,\ldots,t\alpha_n)
\nonumber
\end{eqnarray}
For meson case $n$ = 2.

While the integral over $t$ in Eq. (\ref{eq:mass_int}) is convergent below the threshold $p^2 < (m_{q_1} + m_{q_2})^2$, its convergence above threshold $p^2 \geq (m_{q_1} + m_{q_2})^2$ is guaranteed by augmenting the quark mass by an imaginary part, i.e.  $m_q \rightarrow m_q - i\epsilon, \ \epsilon>0$, in the quark propagator Eq. (\ref{eq:Fock}).  This makes it possible to rotate the integration variable $t$ to the imaginary axis $t\to i t$.  The integral Eq.~(\ref{eq:mass_int}) in turn becomes convergent but obtains an imaginary part corresponding to quark pair production. However, by reducing the scale of integration at the upper limit  corresponding to the introduction of an infrared cutoff
\begin{equation}
\int\limits_0^\infty dt (\ldots) \to \int\limits_0^{1/\lambda^2} dt (\ldots),
\label{eq:conf}
\end{equation}

one can remove all possible thresholds present in the initial quark diagram \cite{Branz:2010}. Thus the infrared cutoff parameter  $\lambda$ effectively guarantees the confinement of quarks within hadrons.

\begin{figure}[h]
\includegraphics{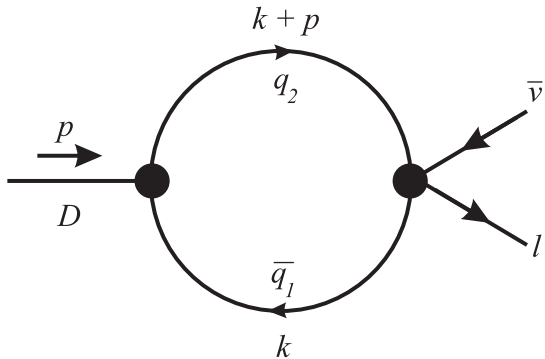}
\caption{Quark model diagrams for the $D$-meson leptonic decay}\label{fig:leptonic_decay}
\end{figure}

Before going for the semileptonic decays, we need to specify the independent model parameters namely size parameter of meson $\Lambda$ and constituent quark masses $m_{q_i}$. These model parameters are determined by fitting calculated decay constants of basic processes such as leptonic (Fig. \ref{fig:leptonic_decay}) and radiative decays to available experimental data or LQCD for vector and pseudoscalar mesons. We use the updated least square fit performed in the recent papers of the model parameters \cite{dubnicka:2016,Gutsche:2015,Ganbold:2014} (all in GeV). We take the infrared cutoff parameter $\lambda$ to be the same throughout this study.
\begin{center}
\begin{tabular}{ccccccccccccc}
$m_{u/d}$ & $m_s$ &  $m_c$ &  $m_b$  & $\lambda$ &\\
\hline
0.241 & 0.428 & 1.67 & 5.05 & 0.181 & GeV
\end{tabular}
\end{center}

\noindent and the size parameters

\begin{center}
\begin{tabular}{ccccccccc}
$\Lambda_D$ & $\Lambda_{D^*}$ & $\Lambda_K$ & $\Lambda_{K^*}$ & $\Lambda_{\pi}$ &\\
\hline
1.6 & 1.53 & 1.01 & 0.80 & 0.87 & GeV
\end{tabular}
\end{center}

We have listed our results for the leptonic decay constants of $D_{(s)}^{(*)}$, $K^{(*)}$ and $\pi$ mesons in the Table \ref{tab:decay_constant}. The decay constants we use in our calculations match quite well with Particle Data Group (PDG), LQCD and QCD sum rules (QCDSR) results.

\begin{table}
\caption{Leptonic decay constants $f_H$ (in MeV)}\label{tab:decay_constant}
\begin{ruledtabular}
\begin{tabular}{cccc}
$f_H$ & Present & Data & Reference\\
\hline
$f_D$   			& 206.1	& 204.6 $\pm$ 5.0 			& PDG \cite{pdg2016}\\
					&			& 207.4 (3.8)						& LQCD \cite{Carrasco:2015}\\
					&			& 210 $\pm$ 11					& QCDSR \cite{Wang:2015}\\
$f_{D^*}$		& 244.3	& 263 $\pm$ 21 				& QCDSR \cite{Wang:2015}\\
					&			& 278 $\pm$ 13 $\pm$ 10 	& LQCD \cite{Becirevic:2012}\\
$f_{D_s}$		& 257.5 & 257.5 $\pm$ 4.6 			& PDG \cite{pdg2016}\\
					&			& 254 (2) (4)						& LQCD \cite{Yang:2015}\\
					&			& 250.2 $\pm$ 3.6				& LQCD \cite{Na:2010}\\
					&			& 247.2 (4.1)						& LQCD \cite{Carrasco:2015}\\
					&			& 259 $\pm$ 10					& QCDSR \cite{Wang:2015}\\
$f_{D_s^*}$	& 272.0 & 308 $\pm$ 21 				& QCDSR \cite{Wang:2015}\\
					&			& 311 $\pm$ 9					& LQCD \cite{Becirevic:2012}\\
$f_{D_s}/f_D$&	1.249& 1.258 $\pm$ 0.038 		& PDG \cite{pdg2016}\\	
					&			& 1.192 (0.22)					& LQCD \cite{Carrasco:2015}\\
					&			& 1.23 $\pm$ 0.07				& QCDSR \cite{Wang:2015}\\
$f_K$			& 156.0	& 155.0 (1.9) 					& LQCD \cite{Carrasco:2015}\\
					&			& 155.37 (34)					& LQCD \cite{Dowdall:2013}\\
					& 			& 157.9 $\pm$ 1.5 			& LQCD \cite{Na:2010}\\
$f_{K^*}$		& 226.8	&	217 $\pm$ 7					& PDG \cite{pdg2016}\\
$f_{\pi}$		& 130.3	& 132.3 $\pm$ 1.6				& LQCD \cite{Na:2010}\\
					&			& 130.39 (20)					& LQCD \cite{Dowdall:2013}
\end{tabular}
\end{ruledtabular}
\end{table}

\section{Form factors}
\label{sec:form_factors}
\begin{figure}[htbp]
\includegraphics[width=0.5\textwidth]{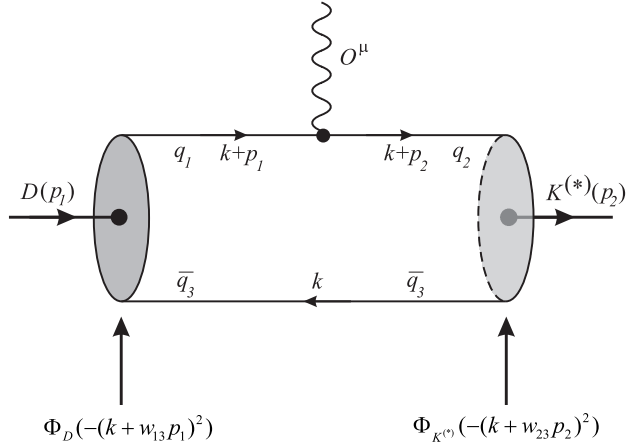}
\caption{Quark model diagrams for the $D$-meson semileptonic decay}
\label{fig:semileptonic}
\end{figure}

In the Standard Model of Particle Physics, semileptonic decays of any meson is caused by weak force in which one lepton and corresponding neutrino is produced in addition to one or more hadrons (Fig. \ref{fig:semileptonic}).

The invariant matrix element for the semileptonic $D \rightarrow K^{(*)} \ell^+ \nu_{\ell}$ decay can be written as
\begin{equation}
M(D \rightarrow K^{(*)} \ell^+ \nu_{\ell}) = \frac{G_F}{\sqrt{2}} V_{cs} \langle K^{(*)} | \bar{s} O^{\mu} c | D \rangle \ \ell^+ O^{\mu} \nu_{\ell}
\end{equation}
where $O^{\mu} = \gamma^\mu(1-\gamma_5)$ is the weak Dirac matrix with left chirality.  The matrix elements for the above semileptonic transitions in the covariant quark model are written as

\begin{widetext}
\begin{eqnarray}\label{eq:ff_DK}
\langle K_{[\bar{d}s]} (p_2) | \bar{s}O^\mu c  | D_{[\bar{d}c]}(p_1) \rangle  &=&
N_c g_D g_K \int \frac{d^4 k}{(2\pi)^4 i} \tilde{\phi}_D (-(k + w_{13} p_1)^2) \tilde{\phi}_K(-(k + w_{23} p_2)^2) \cr && \times \mathrm{tr}[O^\mu S_1(k + p_1) \gamma^5 S_3(k) \gamma^5 S_2(k + p_2)] \cr
&=& F_+(q^2) P^{\mu} +  F_-(q^2) q^{\mu}
\end{eqnarray}
\begin{eqnarray}\label{eq:ff_DK*}
\langle K_{[\bar{d}s]}^* (p_2,\epsilon_{\nu}) | \bar{s}O^\mu c  | D_{[\bar{d}c]}(p_1) \rangle  &=& N_c g_D g_{K^*} \int \frac{d^4 k}{(2\pi)^4 i} \tilde{\phi}_D (-(k + w_{13} p_1)^2) \tilde{\phi}_{K^*}(-(k + w_{23} p_2)^2) \cr && \mathrm{tr}[O^\mu S_1(k + p_1) \gamma^5 S_3(k) \not\!{\epsilon}_{\nu}^\dag S_2(k + p_2)]
\cr & =&\frac{\epsilon_{\nu}^{\dag}}{m_1 + m_2} \left[ -g^{\mu\nu} P\cdot q A_0(q^2) + P^{\mu} P^{\nu}  A_+(q^2) + q^{\mu} P^{\nu}  A_-(q^2) \right. \cr && \left. + i\varepsilon^{\mu\nu\alpha\beta} P_{\alpha} q_{\beta}V(q^2) \right ]
\end{eqnarray}
\end{widetext}

with $P = p_1 + p_2$,  $q = p_1 - p_2$ and $\epsilon_{\nu}$ to be the polarization vector such that $\epsilon_{\nu}^{\dag} \cdot p_2 = 0$ and on-shell conditions of particles require $p_1^2 = m_1^2 = m_D^2$ and $p_2^2 = m_2^2 = m_{K^{(*)}}^2$. Since there are three quarks involved in this transition, we use the notation $w_{ij} = m_{q_j}/(m_{q_i} + m_{q_j})$ ($i$, $j$ = 1, 2, 3) such that $w_{ij} + w_{ji} = 1$.

\section{Numerical Results}
\label{sec:results}
Having determined the necessary model parameters and form factors, we are now in position to present our numerical results. We first compute pure leptonic decays of $D^+$-meson and then using the form factors obtained in Sec. \ref{sec:form_factors}, we compute branching fractions for semileptonic $D$-meson decays.

We compute the pure leptonic decays of $D^+ \to \ell^+ \nu_\ell$ within the Standard Model. The branching fraction for leptonic decay is given by
\begin{equation}
\mathcal{B}(D^+\rightarrow \ell^+ \nu_{\ell}) = \frac{G_F^2}{8 \pi} m_D m_{\ell}^2 \left(1 - \frac{m_{\ell}^2}{m_D^2}\right)^2 f_D^2 |V_{cd}|^2 \tau_D
\end{equation}
where $G_F$ is the fermi coupling constant, $m_D$ and $m_{\ell}$ are the $D$-meson and lepton masses respectively and $\tau_D$ is the $D$-meson lifetime. $f_D$ is the leptonic decay constant of $D$-meson from Table \ref{tab:decay_constant}. The resultant branching fractions for $\ell =\tau, \mu$ and $e$ are given in Table \ref{tab:leptonic_branching}. It is important to note that the helicity flip factor $(1-m_{\ell}^2/m_D^2)$ affects the leptonic branching fractions because of the different lepton masses.
We also compare our results with the experimental data. The branching fraction for $D^+ \to \mu^+\nu_\mu$ shows very good agreement with BESIII \cite{ablikim:2014} and CLEO-c \cite{Eisenstein:2008} data. The branching fractions for $D^+ \to e^+\nu_e$ and $D^+ \to \tau^+\nu_\tau$ also fulfill the experimental constraints.
\begin{table}[htbp]
\caption{Leptonic $D^+$-decay branching fraction ($\tau_{D^+} = 1.040 \times 10^{-12}$ s \cite{pdg2016})}\label{tab:leptonic_branching}
\begin{ruledtabular}
\begin{tabular}{lccc}
Channel & Present & Data & Reference \\
\hline
${{\mathit D^+} \to} \ {{\mathit e}^{+}}{{\mathit \nu}_{{e}}}$				& $8.953 \times 10^{-9}$	& $ < 8.8 \times 10^{-6}$ & PDG \cite{pdg2016}\\
${{\mathit D^+} \to} \ {{\mathit \mu}^{+}}{{\mathit \nu}_{{\mu}}}$		& $3.803 \times 10^{-4}$	& $(3.71 \pm 0.19) \times 10^{-4}$ & BESIII \cite{ablikim:2014}\\
																								& 										& $(3.82 \pm 0.32) \times 10^{-4}$ &  CLEO-c \cite{Eisenstein:2008}\\
${{\mathit D^+} \to} \ {{\mathit \tau}^{+}}{{\mathit \nu}_{{\tau}}} $		& $1.013 \times 10^{-3}$	& $ < 1.2 \times 10^{-3}$ & PDG \cite{pdg2016}
\end{tabular}
\end{ruledtabular}
\end{table}

\begin{figure*}[htbp]
\includegraphics[width=0.45\textwidth]{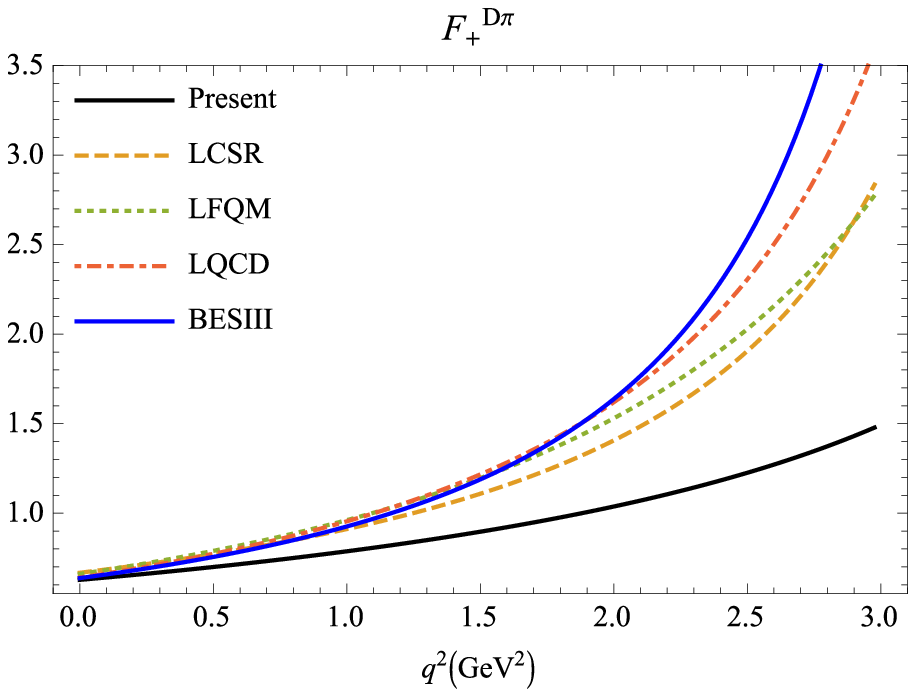}
\hfill\includegraphics[width=0.45\textwidth]{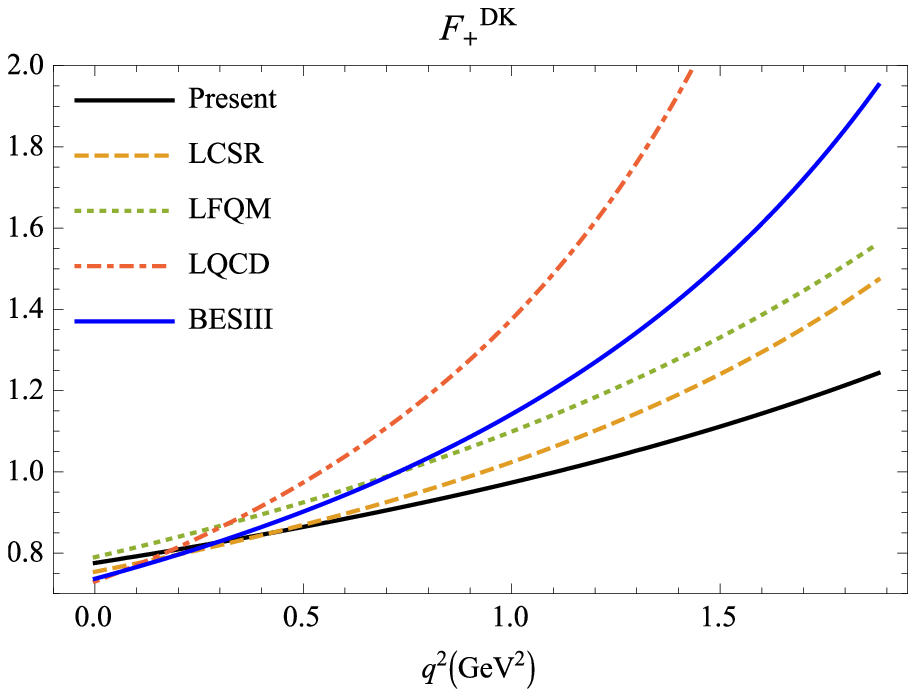}\\
\caption{The results for the form factors appearing in Eq. (\ref{eq:ff_DK}) for semileptonic $D \to \pi$ and $D \to K$ transitions. We compare our plot with the results from LCSR Ref. \cite{Khodjamirian:2009},  LFQM Ref. \cite{Verma:2011}, LQCD Ref. \cite{Aubin:2004} as well with the BESIII data Ref. \cite{ablikim:2015}. }
\label{fig:fplus_dpi}
\end{figure*}

\begin{figure*}[htbp]
\includegraphics[width=0.45\textwidth]{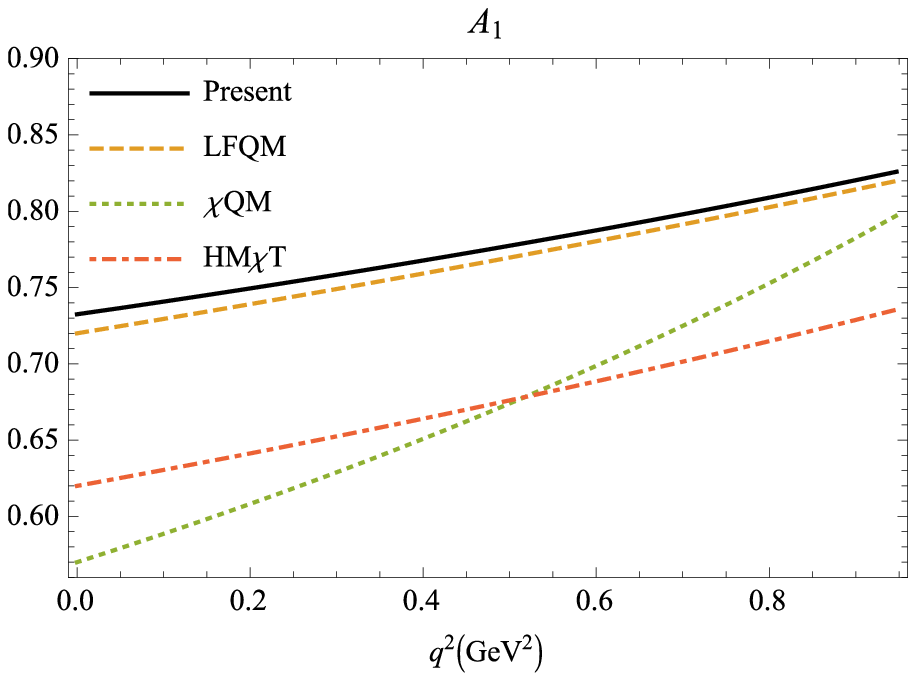}
\hfill\includegraphics[width=0.45\textwidth]{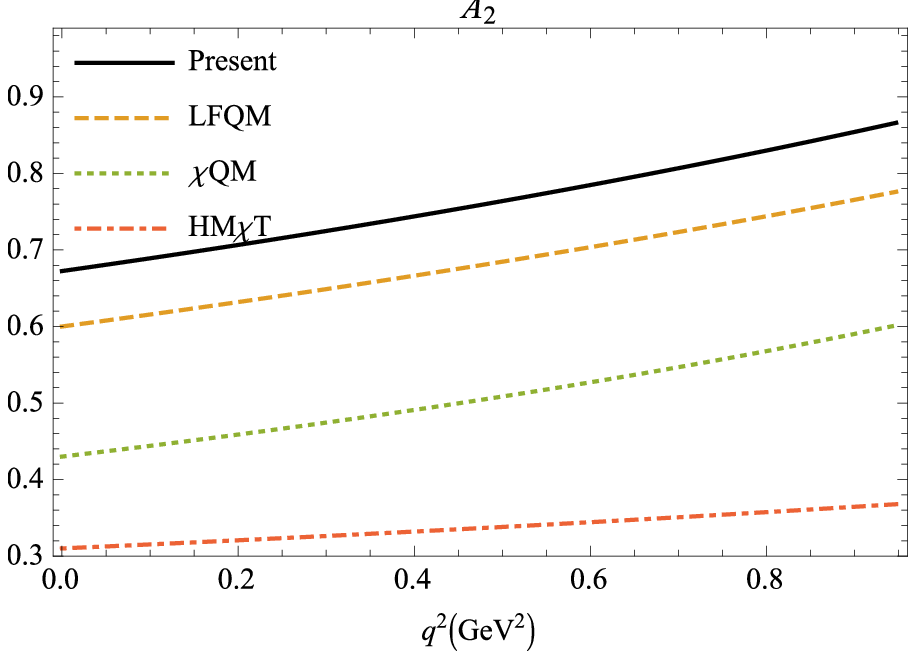}\\
\includegraphics[width=0.45\textwidth]{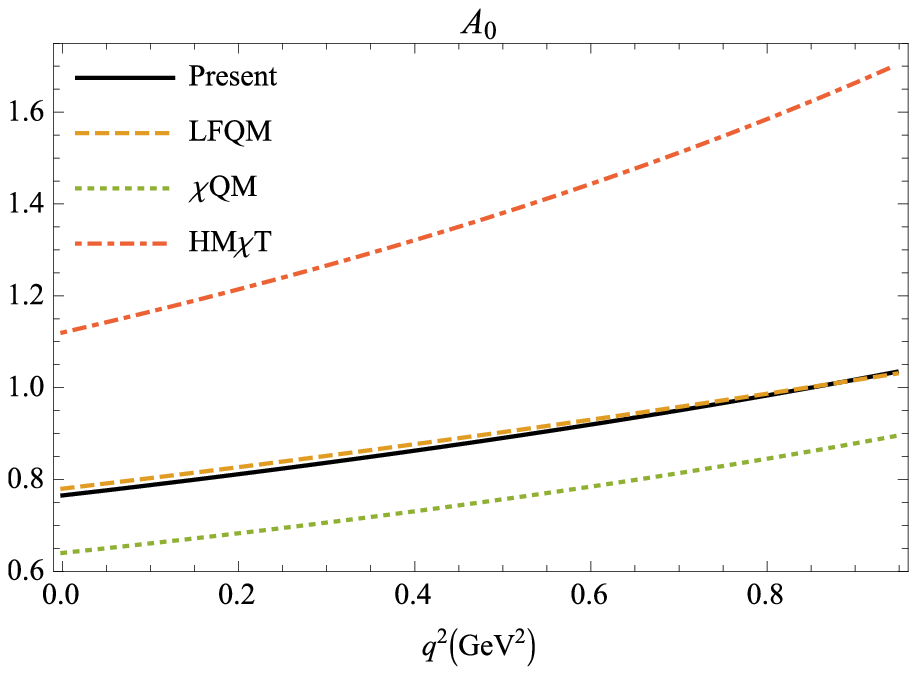}
\hfill\includegraphics[width=0.45\textwidth]{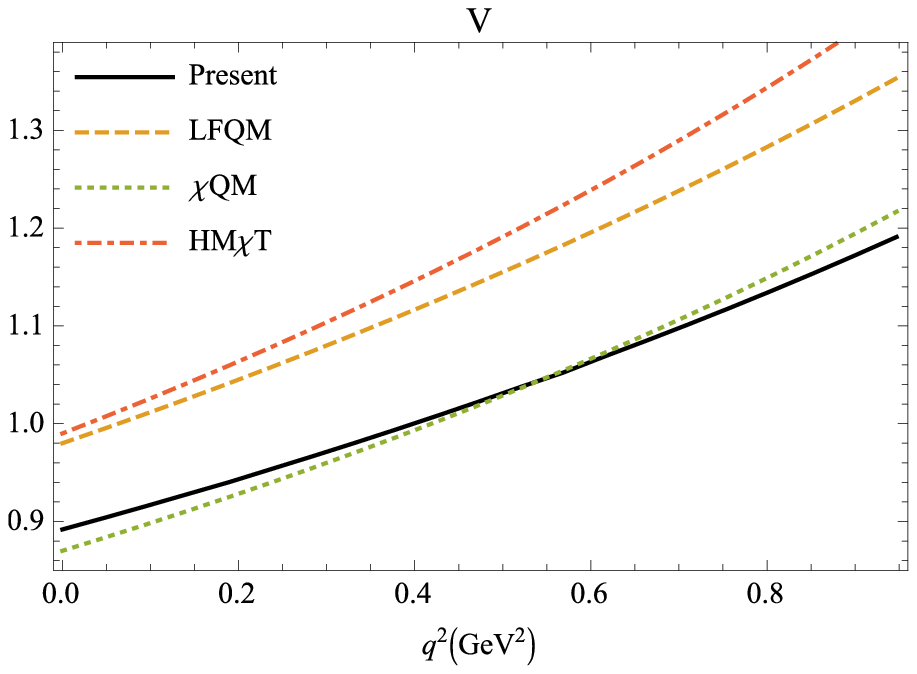}\\
\caption{The form factors appearing in Eq. (\ref{eq:ff_DK*}) for semileptonic $D \to K^*$ transitions. We compare our results with LFQM Ref. \cite{Verma:2011}, chiral quark model ($\chi$QM) Ref. \cite{Palmer:2013} and heavy meson chiral theory (HM$\chi$T) \cite{Fajfer:2005}}
\label{fig:vector_dk}
\end{figure*}

In Figs. \ref{fig:fplus_dpi} and \ref{fig:vector_dk}, we plot our calculated form factors as a function of momentum transfer squared in the entire range $0 \leq q^2 \leq q^2_{max} = (m_D - m_{K^{(*)}})^2$. The multi-dimensional integral (three-fold for semileptonic case) appearing in Eqs. (\ref{eq:ff_DK}) and (\ref{eq:ff_DK*}) are computed numerically using Mathematica. Our form factor results are also well represented by the double-pole parametrization
\begin{equation}
\label{eq:double_pole}
F(q^2) = \frac{F(0)}{1 - a s+ b s^2}, \ \ \ \ s = \frac{q^2}{m_1^2}
\end{equation}

The numerical results of form factors and associated double-pole parameters are listed in Table \ref{tab:double_pole_data}.
\begin{table}[htbp]
\caption{Double pole parameters for the computation of form factors in Eq. (\ref{eq:double_pole})}
\label{tab:double_pole_data}
\begin{ruledtabular}
\begin{tabular}{ccccccc}
				&	$F_+$\ \ & $F_-$\ \ & $A_0$\ \ 	& $A_+$\ \ 	& $A_-$\ \ 	& $V$\ \ \\
\hline
$F(0)$\ \ 	& 0.76\ \ 	& -0.39\ \ 	& 2.07\ \ 		& 0.67\ \ 		& -0.90\ \ 	& 0.89\ \ \\
$a$\ \ 		& 0.72\ \ 	& 0.75\ \ 	& 0.39\ \ 		& 0.84\ \ 		& 0.95\ \ 	& 0.96\ \ 	\\
$b$\ \ 		& 0.046\ \ 	& 0.032\ \ 	& -0.10\ \ 		& 0.087\ \ 		& 0.13\ \ 	& 0.13\ \ 	
\end{tabular}
\end{ruledtabular}
\end{table}
In Fig. \ref{fig:fplus_dpi}, we plot the form factor $F_+$ for $D \to K(\pi) \ell^+ \nu_{\ell}$ decays in the entire kinematical range of momentum transfer. We compare our plot with the results from LCSR Ref. \cite{Khodjamirian:2009},  LFQM Ref. \cite{Verma:2011}, LQCD Ref. \cite{Aubin:2004} as well with the BESIII data Ref. \cite{ablikim:2015}. Our results at maximum recoil point $q^2 \to 0$ are in very good agreement with the other approaches as well as with the experimental result. Similar plot can be obtained for form factor $F_-$. We also plot the vector form factors and for the comparison of the form factors for $D \to K^* \ell^+\nu_{\ell}$ transition with other approaches, we need to write our form factors Eq. (\ref{eq:ff_DK*}) in terms of those used in Ref. \cite{Khodjamirian:2007}. The relations read
\begin{eqnarray}\nonumber
A_0 &=& \frac{m_1 + m_2}{m_1 - m_2}\,A_1 \,, \quad
A_+ = A_2\,,\quad
\end{eqnarray}
\begin{eqnarray}\label{eq:cformfactors}
A_- =  \frac{2m_2(m_1+m_2)}{q^2}\,(A_3 - A_0)\,, \quad
V = V
\end{eqnarray}

The form factors in Eq. (\ref{eq:cformfactors}) also satisfy the constraints
\begin{eqnarray}
\label{eq:constraint}
A_0(0)&=&A_3(0)\\ \nonumber
2m_2A_3(q^2) &=& (m_1+m_2) A_1(q^2) -(m_1-m_2) A_2(q^2)\,.
\end{eqnarray}

Fig. \ref{fig:vector_dk} shows form factors from the present calculation along with the results from  LFQM \cite{Verma:2011}, Chiral Quark Model ($\chi$QM) \cite{Palmer:2013} and with Heavy Meson Chiral Theory (HM$\chi$T) \cite{Fajfer:2005}. The plot shows that our results of the form factors $A_0$, $A_1$ and $A_2$ match with LFQM \cite{Verma:2011} and the vector form factors match with the $\chi$QM \cite{Palmer:2013} where the authors have used energy scaling parameters extracted from modified low energy effective theory in $H\to V$ transitions. Our results show little deviation from those obtained using HM$\chi$T \cite{Fajfer:2005}.
In computation of form factors for $q^2 = 0$ using LCSR, the authors of \cite{Khodjamirian:2009} have used the $\overline{MS}$ scheme for $c$-quark mass and the computation of form factors for $q \geq 0$ is performed in the form of conformal mapping and series parametrization. In the LFQM \cite{Verma:2011}, the authors have used the method of double pole approximation, where as in BESIII \cite{ablikim:2015} and \textit{BABAR} \cite{Lees:2015} experiment, the form factors are parameterized in terms of two and three parameters series expansion respectively.


The differential branching fractions for semileptonic $D \to K \ell^+ \nu_\ell$ decay are computed using \cite{Ivanov:2015,Ivanov:2016}
\begin{eqnarray}\label{eq:dk_diff}
\frac{d\Gamma(D \rightarrow K \ell^+ \nu_{\ell})}{dq^2} &=& \frac{G_F^2 |V_{cs}|^2 |p_2| q^2 v^2}{12 (2\pi)^3 m_1^2} \nn & \times & \Big((1+\delta_{\ell}) \mathcal{H}_L + 3 \delta_{\ell} \mathcal{H}_{SL}\Big)
\end{eqnarray}

where the helicity flip factor $\delta_{\ell} = m_{\ell}^2/2 q^2$,  $|p_2| = \lambda^2(m_1^2,m_2^2,q^2)/2 m_1$ is momentum of $K$ meson in the rest frame of $D$-meson  and velocity-type parameter $v = 1 - m_{\ell}^2/q^2$.

The bilinear combinations of the helicity amplitudes $\mathcal{H}$ are defined as \cite{Faessler:2002},
\begin{equation}
\mathcal{H}_L = |H_0|^2,\ \ \mathcal{H}_S = |H_t|^2,  \ \ \mathcal{H}_{SL}=Re(H_0 H_t^{\dag})
\end{equation}
and the helicity amplitudes are expressed via the form factor in the matrix element as,
\begin{equation}
H_t = \frac{1}{\sqrt{q^2}} (PqF_+ + q^2F_-)
\end{equation}

\begin{equation}
H_0 = \frac{2 m_1 |p_2|}{\sqrt{q^2}} F_+
\end{equation}

Similarly the differential branching fractions for semileptonic $D \to K^* \ell^+ \nu_\ell$ decay is computed by \cite{Ivanov:2015,Ivanov:2016}
\begin{eqnarray}\label{eq:dks_diff}
\frac{d\Gamma(D \rightarrow K^* \ell^+ \nu_{\ell})}{dq^2} &=& \frac{G_F^2 |V_{cs}|^2 |p_2| q^2 v^2}{12 (2\pi)^3 m_1^2} \\ & \times & \Big((1+\delta_{\ell}) (\mathcal{H}_U + \mathcal{H}_L) + 3 \delta_{\ell} \mathcal{H}_{S}\Big)
\nonumber\end{eqnarray}

\begin{figure*}[htbp]
\includegraphics[width=0.45\textwidth]{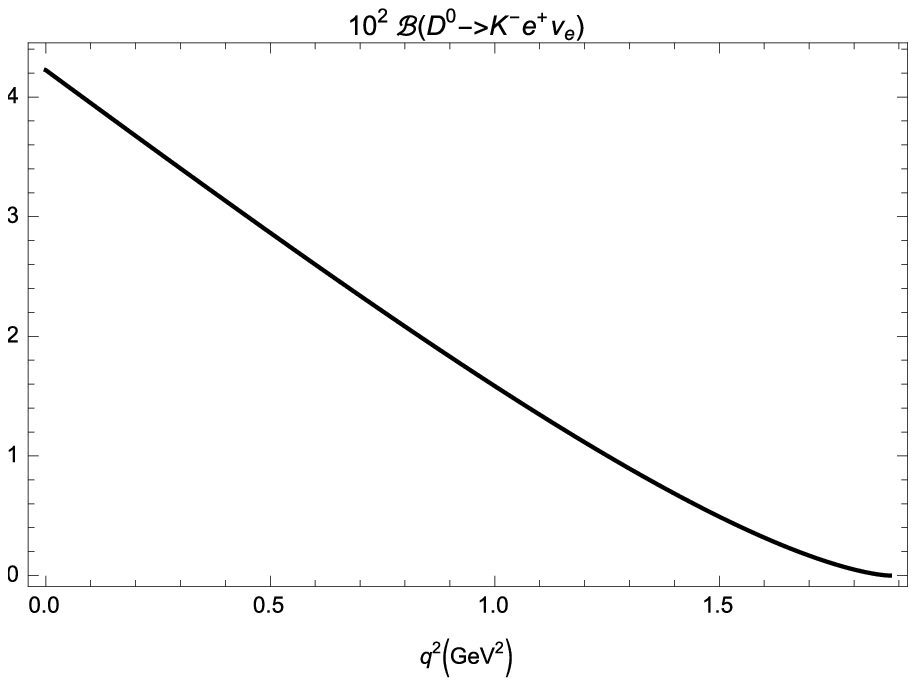}
\hfill\includegraphics[width=0.45\textwidth]{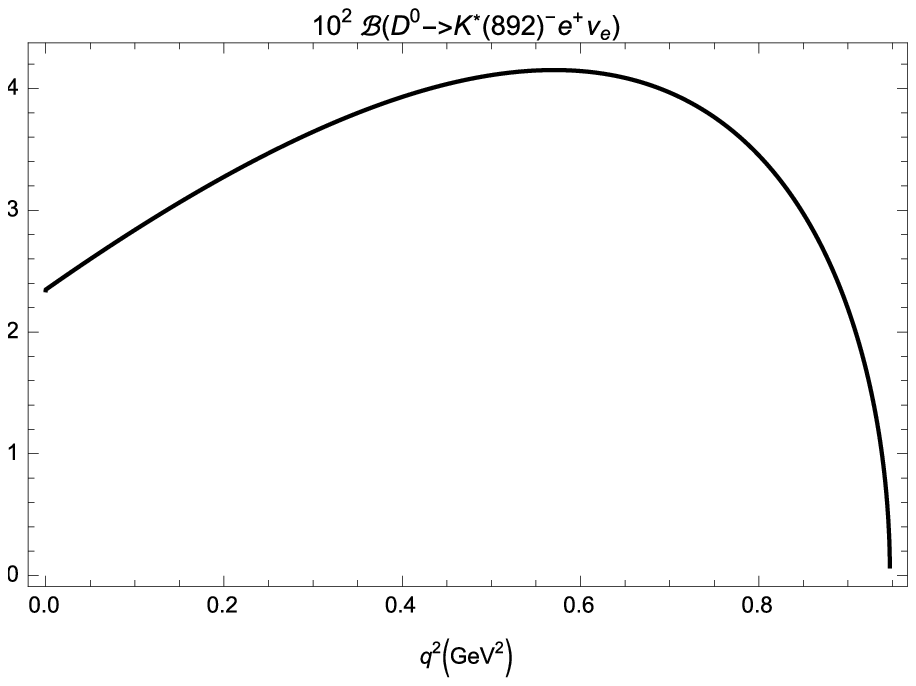}\\
\includegraphics[width=0.45\textwidth]{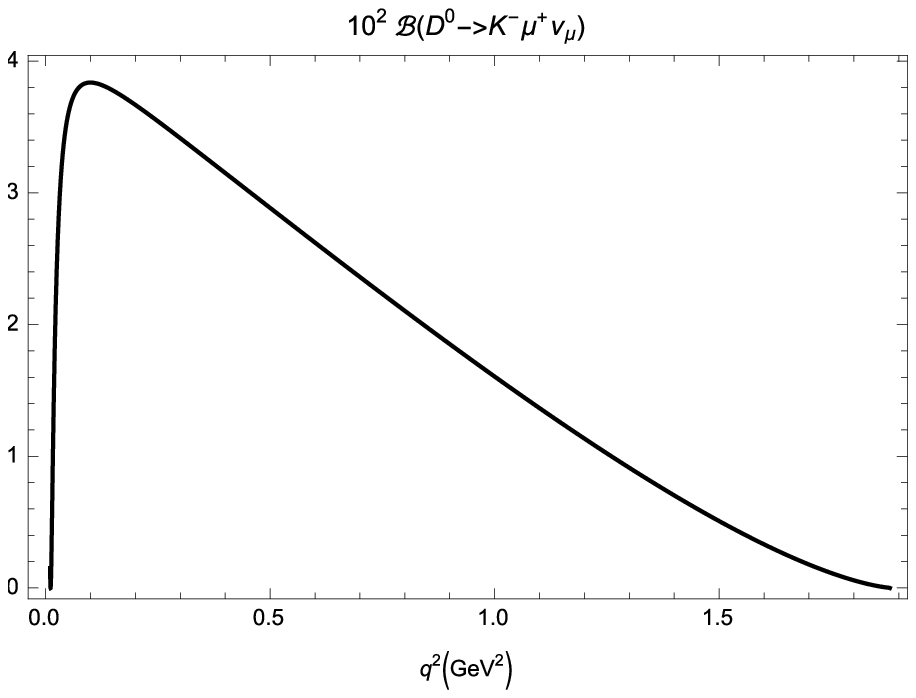}
\hfill\includegraphics[width=0.45\textwidth]{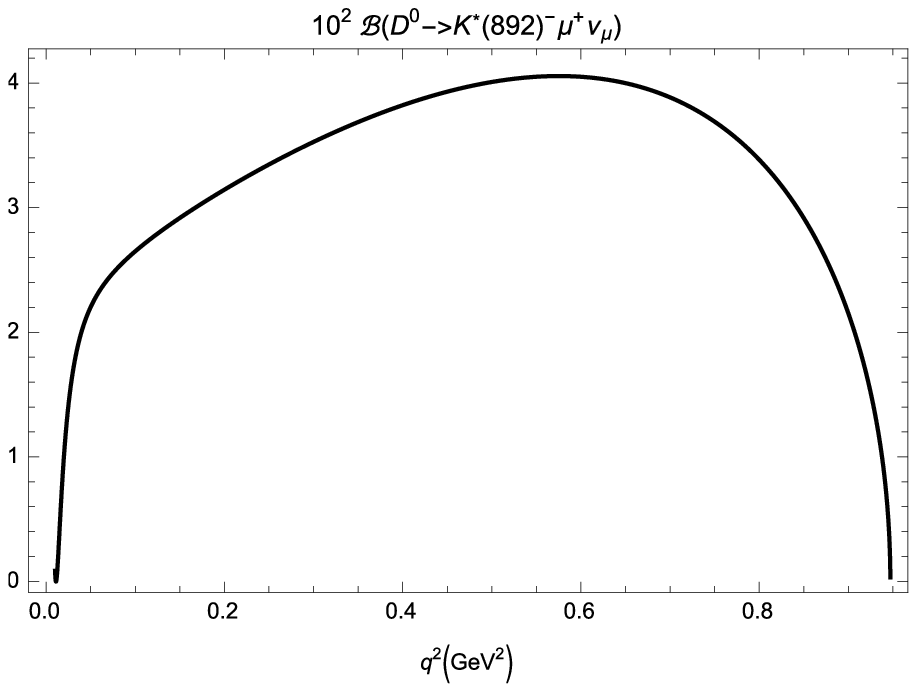}\\
\caption{Differential branching fractions of the decays $D \to K^{(*)} {\mathit{\ell}}^+ \nu_{\ell}$}
\label{fig:branching}
\end{figure*}

The bilinear combinations of the helicity amplitudes $\mathcal{H}$ are defined as  \cite{Faessler:2002}
\begin{equation} \nonumber
\mathcal{H}_U = |H_{+1+1}|^2 + |H_{-1-1}|^2, \ \ \mathcal{H}_P = |H_{+1+1}|^2 - |H_{-1-1}|^2,
\end{equation}

\begin{equation}
\mathcal{H}_L = |H_{00}|^2, \  \ \mathcal{H}_S = |H_{t0}|^2, \ \ \mathcal{H}_{SL} = Re(H_{00} H_{t0}^{\dag})
\end{equation}
here also the helicity amplitudes are expressed via the form factor in the matrix element as
\begin{equation}
H_{t0} = \frac{1}{m_1 + m_2} \frac{m_1 |p_2|}{m_2 \sqrt{q^2}} (Pq (-A_0 + A_+) + q^2 A_-)
\end{equation}
\begin{equation}
H_{\pm 1 \pm 1} = \frac{1}{m_1 + m_2} (-Pq A_0 \pm 2 m_1 |p_2| V)
\end{equation}
\begin{eqnarray}
H_{00} &=& \frac{1}{m_1 + m_2} \frac{1}{2 m_2 \sqrt{q^2}} \nn
& \times & (-Pq (m_1^2 - m_2^2 - q^2) A_0 + 4 m_1^2 |p_2|^2 A_+)
\end{eqnarray}
\begin{table*}[htbp]
\begin{center}
\caption{Branching fractions of $D \to K^{(*)} {\mathit{\ell}}^+ \nu_{\ell}$ and $D \to \pi \ell^+ \nu_{\ell}$ (in \%)}\label{tab:results}
\begin{ruledtabular}
\begin{tabular}{l ccc}
\multicolumn{1}{c}{Channel} & Present & Data & \ \ Reference\\
\hline
${\mathit D}^+ \rightarrow {{\overline{\mathit K}}^{0}}{{\mathit e}^{+}}{{\mathit \nu}_{{e}}}$	& 8.84 	& 8.60 $\pm$ 0.06 $\pm$ 0.15 		& \ \ BESIII \cite{ablikim:2017lks}\\
																			& 			&8.83 $\pm$ 0.10 $\pm$ 0.20 		& \ \ CLEO-c \cite{Besson:2009}\\
${\mathit D}^+ \rightarrow {{\overline{\mathit K}}^{0}}{{\mathit \mu}^{+}}{{\mathit \nu}_{{\mu}}}$	& 8.60	& 8.72 $\pm$ 0.07 $\pm$ 0.18 		& \ \ BESIII \cite{ablikim:2016sqt}\\
${\mathit D}^+ \rightarrow {{\mathit \pi}^{0}}{{\mathit e}^{+}}{{\mathit \nu}_{{e}}}$		& 0.619 & 0.363 $\pm$ 0.08 $\pm$ 0.05 	& \ \ BESIII \cite{ablikim:2017lks}\\
																			& 			& 0.405 $\pm$ 0.016 $\pm$ 0.009 	& \ \ CLEO-c \cite{Besson:2009}\\
${\mathit D}^+ \rightarrow {{\mathit \pi}^{0}}{{\mathit \mu}^{+}}{{\mathit \nu}_{{\mu}}}$		& 0.607	 & -- & --\\	
${{\mathit D}^{+}} \rightarrow {{\overline{\mathit K}}^{*}{(892)}^{0}}{{\mathit e}^{+}}{{\mathit \nu}_{{e}}}$		& 8.35	 & -- & --\\
${{\mathit D}^{+}} \rightarrow {{\overline{\mathit K}}^{*}{(892)}^{0}}{{\mathit \mu}^{+}}{{\mathit \nu}_{{\mu}}}$	& 7.94	 & -- & --\\
${{\mathit D}^{0}} \rightarrow {{\mathit K}^{-}}{{\mathit e}^{+}}{{\mathit \nu}_{{e}}}$ & 3.46	& 3.538 $\pm$ 0.033 & \ \ PDG \cite{pdg2016}\\
																			& 			& 3.505 $\pm$ 0.014 $\pm$ 0.033 & \ \ BESIII \cite{ablikim:2015}\\
																			& 			& 3.50 $\pm$ 0.03 $\pm$ 0.04 		& \ \ CLEO-c \cite{Besson:2009}\\
																			&			& 3.45 $\pm$ 0.07 $\pm$ 0.20		& \ \ Belle \cite{Widhalm:2006}\\
${{\mathit D}^{0}} \rightarrow {{\mathit K}^{-}}{{\mathit \mu}^{+}}{{\mathit \nu}_{{\mu}}}$		& 3.36 & 3.33 $\pm$ 0.13 &	\ \ PDG \cite{pdg2016}	\\
																																	& 		& 3.505 $\pm$ 0.014 $\pm$ 0.033 	& \ \ BESIII \\
${{\mathit D}^{0}} \rightarrow {{\mathit \pi}^{-}}{{\mathit e}^{+}}{{\mathit \nu}_{{e}}}$		& 0.239 & 0.2770 $\pm$ 0.0068 $\pm$ 0.0092& \ \ \textit{BABAR} \cite{Lees:2015}\\
																			& 			& 0.295 $\pm$ 0.004 $\pm$ 0.003 	& \ \ BESIII \cite{ablikim:2015}\\
																			& 			& 0.288 $\pm$ 0.008 $\pm$ 0.003 	& \ \ CLEO-c \cite{Besson:2009}\\
																			& 			& 0.255 $\pm$ 0.019 $\pm$ 0.016 & \ \ Belle \cite{Widhalm:2006}\\
${{\mathit D}^{0}} \rightarrow {{\mathit \pi}^{-}}{{\mathit \mu}^{+}}{{\mathit \nu}_{{\mu}}}$			& 0.235 & 0.238 $\pm$ 0.024						& \ \ PDG \cite{pdg2016}\\
${{\mathit D}^{0}} \rightarrow {{\mathit K}^{*}{(892)}^{-}}{{\mathit e}^{+}}{{\mathit \nu}_{{e}}}$& 3.25 & 2.16 $\pm$ 0.16 & \ \ PDG \cite{pdg2016}  \\
${{\mathit D}^{0}} \rightarrow {{\mathit K}^{*}{(892)}^{-}}{{\mathit \mu}^{+}}{{\mathit \nu}_{{\mu}}}$	& 3.09 	& 1.92 $\pm$ 0.25 & \ \ PDG \cite{pdg2016}
\end{tabular}
\end{ruledtabular}
\end{center}
\end{table*}
In Fig. \ref{fig:branching}, we present our results for differential branching fractions of $D \to K^{(*)} \ell^+ \nu_\ell$ in the entire kinematical range of momentum transfer. The semileptonic branching fractions in Eqs. (\ref{eq:dk_diff}) and (\ref{eq:dks_diff}) are computed by numerically integrating the differential branching fractions shown in Fig. \ref{fig:branching}. The branching fractions for $D \to K^{(*)} \ell^+\nu_{\ell}$ and $D \to \pi \ell^+\nu_{\ell}$ are presented in Table \ref{tab:results}. We also compare our results with experimental results. The results for $\mathcal{B} ({\mathit D}^+ \rightarrow {{\overline{\mathit K}}^{0}}{{\mathit \ell}^{+}}{{\mathit \nu}_{{\ell}}})$ and $\mathcal{B} ({{\mathit D}^{0}} \rightarrow {{\mathit K}^{-}}{{\mathit \ell}^{+}}{{\mathit \nu}_{{\ell}}})$, ($\ell = e$ and $\mu$) show excellent agreement with the recent BESIII data \cite{ablikim:2017lks,ablikim:2016sqt,ablikim:2015} as well with the other experimental collaborations. Also the ratios of the different semileptonic decay widths for the channels $D \to K \ell^+ \nu_{\ell}$ are presented in Table \ref{tab:iso_spin} and our results are well within the isospin conservation rules given in Ref. \cite{Korner:1989}. We also present our results for $\mathcal{B}({{\mathit D}^{0}} \rightarrow {{\mathit K}^{*}{(892)}^{-}}{{\mathit \ell}^{+}}{{\mathit \nu}_{{e}}})$ but our results overestimate the data given in PDG \cite{pdg2016}. This deviation of the present study within the Standard Model might be explained through hadronic uncertainty or ratios of differential distributions for longitudinal and transverse polarizations of these $K^*$ mesons \cite{Fajfer:2015}. The FOCUS \cite{Link:2002} and CLEO-c \cite{Shepherd:2006} experiments have also reported mixing of scalar amplitudes with dominant vector decays. 
These observations open up new possibilities of investigations in charm semileptonic decays. There have also been attempts to explain these exclusive decays using $R$-parity violating supersymmetric effects \cite{Wang:2014} and their direct correlation with possible supersymmetric signals expected from LHC and BESIII data. We predict the branching fractions for ${{\mathit D}^{+}} \rightarrow {{\overline{\mathit K}}^{*}{(892)}^{0}}{{\mathit \ell}^{+}}{{\mathit \nu}_{{e}}}$ but we do not compare our results since no experimental results available for this channel.

We also present our results for branching fractions of $ {\mathit D}^+ \rightarrow {{\mathit \pi}^{0}}{{\mathit \ell}^{+}}{{\mathit \nu}_{{\ell}}}$ and $ {{\mathit D}^{0}} \rightarrow {{\mathit \pi}^{-}}{{\mathit \ell}^{+}}{{\mathit \nu}_{{\ell}}}$ transitions. Our prediction for $\mathcal{B} ({\mathit D}^+ \rightarrow {{\mathit \pi}^{0}}{{\mathit e}^{+}}{{\mathit \nu}_{e}})$ is higher than BESIII \cite{ablikim:2017lks} and CLEO-c  data \cite{Besson:2009} while the trend is opposite in the case of $\mathcal{B} ({{\mathit D}^{0}} \rightarrow {{\mathit \pi}^{-}}{{\mathit e}^{+}}{{\mathit \nu}_{{e}}})$. The deviation of the $\mathcal{B} ({\mathit D}^+ \rightarrow {{\mathit \pi}^{0}}{{\mathit e}^{+}}{{\mathit \nu}_{e}})$ from experimental and LQCD data might be attributed to the computed form factors. However, our $\mathcal{B} ({\mathit D}^0 \rightarrow {{\mathit \pi}^{-}}{{\mathit e}^{+}}{{\mathit \nu}_{e}})$ is in close proximity to that by Belle \cite{Widhalm:2006} and $\mathcal{B} ({\mathit D}^0 \rightarrow {{\mathit \pi}^{-}}{{\mathit \mu}^{+}}{{\mathit \nu}_{\mu}})$ is in excellent agreement with PDG data \cite{pdg2016}.

\begin{figure*}[htbp]
\includegraphics[width=0.45\textwidth]{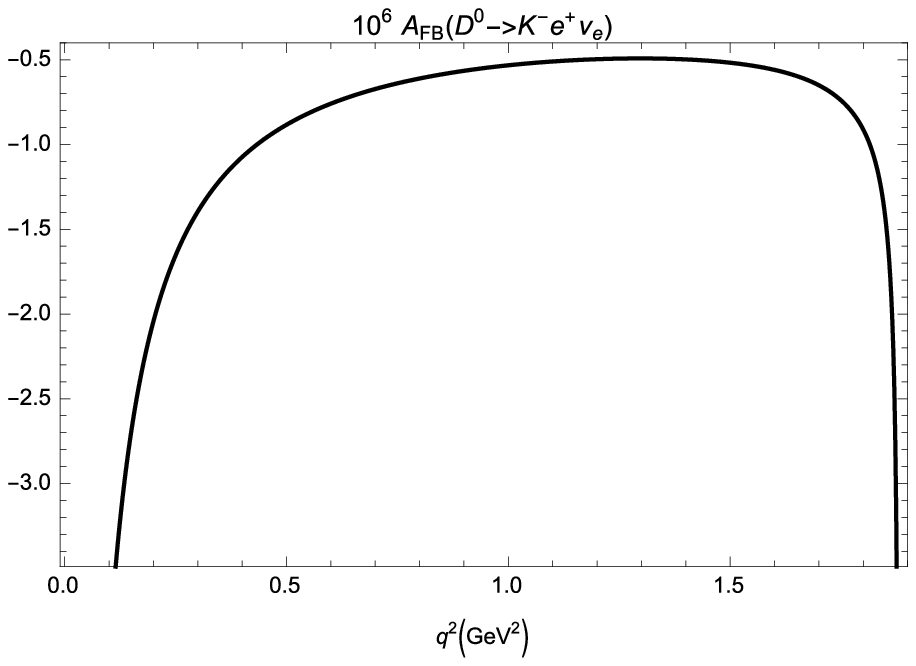}
\hfill\includegraphics[width=0.45\textwidth]{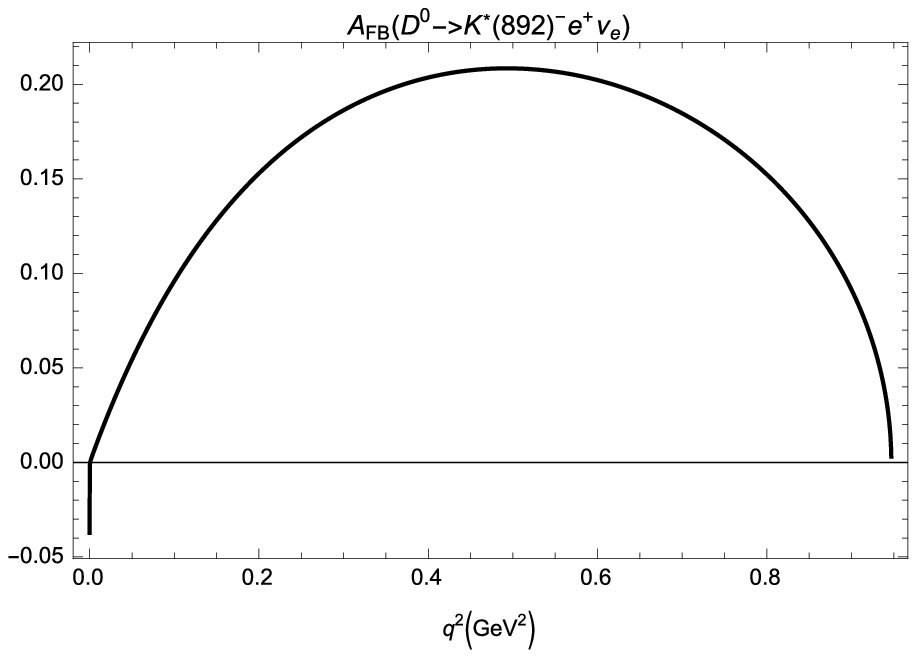}\\
\includegraphics[width=0.45\textwidth]{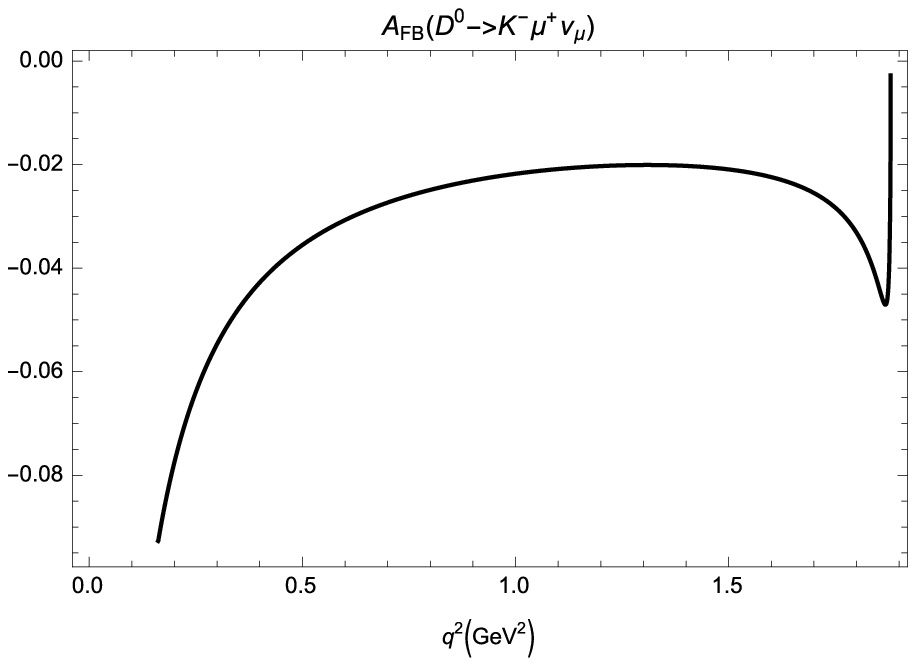}
\hfill\includegraphics[width=0.45\textwidth]{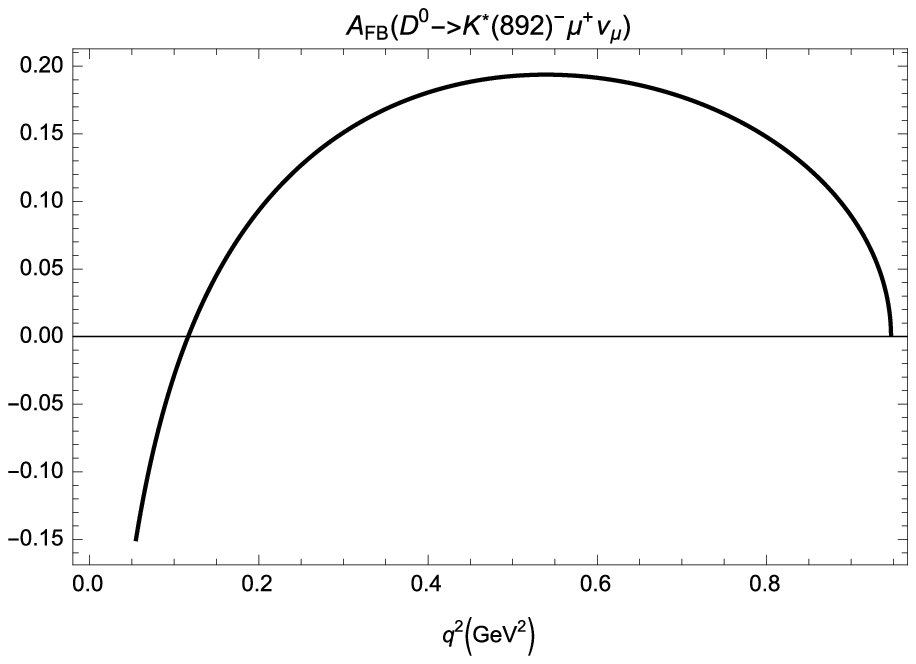}\\
\caption{Forward-backward asymmetries of the decays $D \to K^{(*)} \ell^+ \nu_{\ell}$}
\label{fig:fb}
\end{figure*}

\begin{table}[htbp]
\begin{ruledtabular}
\caption{Ratios of the semileptonic decays of $D$ mesons}\label{tab:iso_spin}
\begin{tabular}{c  c}
Ratio & \ \ Value\\
\hline
$\Gamma ({{\mathit D}^{0}} \rightarrow {{\mathit K}^{-}}{{\mathit e}^{+}}{{\mathit \nu}_{{e}}})/\Gamma({{\mathit D}^{+}} \rightarrow {{\overline{\mathit K}}^{0}}{{\mathit e}^{+}}{{\mathit \nu}_{{e}}})$ & \ \ 1.02\\
$\Gamma({{\mathit D}^{0}} \rightarrow {{\mathit K}^{-}}{{\mathit \mu}^{+}}{{\mathit \nu}_{{\mu}}})/\Gamma({{\mathit D}^{+}} \rightarrow {{\overline{\mathit K}}^{0}}{{\mathit \mu}^{+}}{{\mathit \nu}_{{\mu}}})$ & \ \ 0.99\\
$\Gamma({{\mathit D}^{+}} \rightarrow {{\overline{\mathit K}}^{0}}{{\mathit \mu}^{+}}{{\mathit \nu}_{{\mu}}}) / \Gamma({{\mathit D}^{+}} \rightarrow {{\overline{\mathit K}}^{0}}{{\mathit e}^{+}}{{\mathit \nu}_{{e}}}) $ & \ \ 0.97
\end{tabular}
\end{ruledtabular}
\end{table}

We also list some more physical observables in terms of helicity amplitudes. We have already shown the computed differential branching fractions in Fig. \ref{fig:branching}.
Next, the helicity amplitudes defined above are used to plot the forward-backward asymmetry in Fig. \ref{fig:fb} for $D \to K^{(*)} \ell^+ \nu_\ell$ in the entire kinematical range of momentum transfer.
We use the following relation for plotting the forward-backward asymmetry ($A_{FB}$) \cite{Gutsche:2015,Ivanov:2015}
\begin{equation}
\mathcal{A}_{FB} (q^2) = -\frac{3}{4} \frac{\mathcal{H}_P + 4 \delta_{\ell} \mathcal{H}_{SL}}{(1+\delta_{\ell}) (\mathcal{H}_U + \mathcal{H}_L) + 3 \delta_{\ell} \mathcal{H}_{S}}
\end{equation}
It is evident from Fig. \ref{fig:fb} that the $A_{FB} (q^2)$ for $D \to K \ell^+ \nu_\ell$ and $D \to K^* \ell^+ \nu_\ell$ are similar for both $e$ and $\mu$ modes. $A_{FB} (q^2) \to 0$ for in the both zero recoil and larger recoil limits because of the zero recoil relations of the helicity functions $\mathcal{H}_P = \mathcal{H}_{SL} = 0$ and longitudinal dominance in the partial rates at the maximum recoil.

Also the lepton and hadron side convexity parameter are defined as \cite{Gutsche:2015,Ivanov:2015}
\begin{equation}\label{eq:convex1}
C_F^\ell = \frac{3}{4} \frac{(1-2 \delta_\ell)(\mathcal{H}_U - 2 \mathcal{H}_L)}{(1+\delta_{\ell}) (\mathcal{H}_U + \mathcal{H}_L) + 3 \delta_{\ell} \mathcal{H}_{S}}
\end{equation}
and
\begin{equation}\label{eq:convex2}
C_F^h = -\frac{3}{2} \frac{(1+\delta_{\ell}) (\mathcal{H}_U - 2 \mathcal{H}_L) - 6 \delta_\ell \mathcal{H}_S}{(1+\delta_{\ell}) (\mathcal{H}_U + \mathcal{H}_L) + 3 \delta_{\ell} \mathcal{H}_{S}}
\end{equation}

\begin{table}[htbp]
\begin{ruledtabular}
\caption{Averages of forward-backward asymmetry and convexity parameters}\label{tab:average}
\begin{tabular}{c  c  c  c c}
Channel  & $\ell$ & $\langle A_{FB}^\ell \rangle$ & $\langle C_F^\ell \rangle$ & $\langle C_F^h \rangle$\\
\hline
$D \to K$		& $e$ 		& -4.27 $\times$ 10$^{-6}$	& -1.5	& 3\\
					& $\mu$	& -0.058							& -1.32	& 3\\
\hline
$D \to K^*$	& $e$ 		& 0.17	& -0.45	 & 0.91\\
					& $\mu$	& 0.13	& -0.37	 & 0.89
\end{tabular}
\end{ruledtabular}
\end{table}

The plot for the convexity parameters Eqs. (\ref{eq:convex1}) and (\ref{eq:convex2}) as a function of entire momentum transfer range can easily be obtained. In Table \ref{tab:average}, we give the $q^2$ averages of the above observables. Note that in order to obtain the averages of these observables, we need to multiply the numerator and denominator by phase space factor $|p_2|q^2v^2$. Also in computation of leptonic and semileptonic branching fractions, forward-backward asymmetry and convexity parameters, the values of CKM matrices namely $|V_{cs}|$ and $|V_{cd}|$, meson masses, lepton masses and their lifetimes are taken from PDG \cite{pdg2016}.

\section{Conclusion}
In this article, we have analysed the leptonic ($D^+ \to {{\mathit e}^{+}}{{\mathit \nu}_{{e}}}$) and semileptonic ($D \to K^{(*)} \ell^+ \nu_{\ell}$,  $D \to \pi \ell^+ \nu_{\ell}$) decays using covariant quark model with infrared confinement within the standard model framework. The ratios of the partial widths are found to be consistent with the isospin conservation holding within uncertainties in experimental data. It is interesting to note here that the $\mathcal{B} ({\mathit D}^+ \rightarrow {{\mathit \pi}^{0}}{{\mathit \ell}^{+}}{{\mathit \nu}_{{\ell}}})$ deviate from existing data while $\mathcal{B} ({{\mathit D}^{0}} \rightarrow {{\mathit \pi}^{-}}{{\mathit \ell}^{+}}{{\mathit \nu}_{{\ell}}})$ match well. Further exploration to this observation may lead to interesting outcome.

The deviation of branching fractions in case of $D\to K^*\ell\nu_{\ell}$ might be understood by underlying hadronic uncertainty or ratios of differential distributions for longitudinal and transverse polarizations of the $K^*$ mesons. We are looking forward to analyse $D \to K^* \ell^+ \nu_\ell$ decay and expect the experimental facilities to throw more light on their form factor shapes in forthcoming attempts that will help in understanding the charm decays and possibly the dynamics of these systems beyond the standard model.
\label{sec:summary}
\section*{Acknowledgment}
We thank Prof. Mikhail A. Ivanov for the continuous support through out this work and providing critical remarks for improvement of the manuscript. NRS would like to thank Bogoliubov Laboratory of Theoretial Physics, Joint Institute for Nuclear Research for warm hospitality during Helmholtz-DIAS International Summer School ``Quantum Field Theory at the Limits: from Strong Field to Heavy Quarks" where this work was initiated. This work is done under Major Research Project F.No.42-775/2013(SR) with financial support from the University Grants Commission of India.

\end{document}